\documentstyle[12pt,epsfig]{article}
\newcommand{\be}{\begin{equation}}
\newcommand{\ee}{\end{equation}}
\begin{document}  
\topmargin 0pt
\oddsidemargin=-0.4truecm
\evensidemargin=-0.4truecm
\renewcommand{\thefootnote}{\fnsymbol{footnote}}
\newpage
\setcounter{page}{0}
\begin{titlepage}   
\vspace*{-2.0cm}  
%%%\vspace*{-1.0cm}
\begin{flushright}
%FISIST/01-01/CFIF \\
%%\vspace*{-0.2cm}
hep-ph/0112104
\end{flushright}
\vspace*{0.5cm}
\begin{center}
{\Large \bf Solar Neutrinos with Magnetic Moment: Rates and Global Analysis} 
\vspace{1.0cm}

{\large 
Jo\~{a}o Pulido
\footnote{E-mail: pulido@cfif.ist.utl.pt}}\\  
{\em Centro de F\'\i sica das Interac\c c\~oes Fundamentais (CFIF)} \\
{\em Departamento de Fisica, Instituto Superior T\'ecnico }\\
{\em Av. Rovisco Pais, P-1049-001 Lisboa, Portugal}\\
\end{center}
\vglue 0.8truecm

\begin{abstract}
A statistical analysis of the solar neutrino data is presented assuming 
the solar neutrino deficit to be resolved by the resonant interaction of the neutrino 
magnetic moment with the solar magnetic field. Four field profiles are
investigated, all exhibiting a rapid increase across the bottom of the convective
zone, one of them closely following the requirements from recent solar physics
investigations. First a 'rates only' analysis is performed whose best fits
appear to be remarkably better than all fits from oscillations. A global
analysis then follows with the corresponding best fits of a comparable
quality to the LMA one. Despite the fact that the resonant spin flavour
precession does not predict any day/night effect, the separate SuperKamiokande
day and night data are included in the analysis in order to allow for a direct
comparison with oscillation scenarios. Remarkably enough, the best fit for
rates and global analysis which is compatible with most astrophysical bounds
on the neutrino magnetic moment is obtained from the profile which most closely
follows solar physics requirements. Allowing for a peak field value of $3\times
10^5G$, it is found in this case that $\Delta m^2_{21}=1.45\times10^{-8}eV^2$,
$\mu_{\nu}=3.2\times10^{-12}\mu_{B}$ (65\%CL). The new forthcoming experiments 
on solar neutrino physics (Kamland and Borexino) will be essential to ascertain 
whether this fact is incidental or essential. 
  
\end{abstract}
\end{titlepage}   

%\vspace {30mm}
%\newpage
\section{Introduction}

Whereas the apparent anticorrelation of the neutrino event rate with sunspot 
activity claimed long ago by the Homestake collaboration \cite{Homestake} 
remained unconfirmed by other experiments \cite{SuperK}, \cite{SAGE}, 
\cite{Gallex+GNO} and theoretical analyses \cite{Walther}, the magnetic moment 
solution to the solar neutrino problem is at present an important possibility 
to be explored in the quest for an explanation of the solar neutrino 
deficit. This is the idea, originally proposed by Cisneros \cite{Cisneros},
and later revived by Voloshin, Vysotsky and Okun \cite{VVO},  
that a large magnetic moment of the neutrino may interact with the magnetic 
field of the sun, converting weakly active to sterile neutrinos.
It now appears in fact that this deficit is energy dependent, in the sense
that neutrinos of different energies are suppressed differently. In order to 
provide an energy dependent deficit, the conversion mechanism from active to 
nonactive neutrinos must be resonant, with the location of the critical density 
being determined by the neutrino energy. Thus was developed the idea of the
resonance spin flavour precession (RSFP) proposed in 1988 \cite{LMA}. It involves
the simultaneous flip of both chirality and flavour consisting basically in the 
assumption that the neutrino conversion due to
magnetic moment and magnetic field takes place through a resonance inside matter 
in much the same way as matter oscillations \cite{MSW}. 
A sunspot activity related event rate 
in a particular experiment would hence imply that most of the neutrinos with energies 
relevant to that experiment have their resonances in the sunspot range. However,
the depth of sunspots is unknown (they may not extend deeper than a few hundred
kilometers) and the observed field intensity is too small in sunspots to allow
for a significant conversion. The anticorrelation argument has therefore lost 
its appeal for several years now.

Despite the absence of the anticorrelation argument,
several main reasons may be invoked to motivate RSFP and investigating its consequences
for solar neutrinos. In fact both RSFP and all oscillation scenarios indicate a 
drop in the survival probability from the low energy (pp) to the 
intermediate energy neutrino sector ($^7$Be, CNO, pep) and a subsequent moderate rise as 
the energy increases further into the $^8$B sector. The magnetic field profiles
providing good fits to the event rates from solar neutrino experiments typically show 
the characteristic of a sharp rise in intensity at some point in the solar 
interior, followed by a progressive moderate decrease \cite{PA}, \cite{Valle}. 
This is in opposite correspondence with the energy dependence of the probability  
in the sense that the strongest field intensities correspond to the smallest survival 
probabilities. Hence RSFP offers a very suggestive
explanation for the general shape of the probability, which naturally appears
as a consequence of the field profile. On the other hand, from solar physics and 
helioseismology such a sharp rise and peak field intensity is expected to occur along 
the tachoclyne, the region extending from the upper radiative zone to the lower 
convective zone, where the gradient of the angular velocity of solar matter is 
different from zero \cite{Parker}, \cite{ACT}. Furthermore, it has become 
clear \cite{PA,Valle,GN} that RSFP provides event rate fits 
from the solar neutrino experiments that are remarkably better than oscillation ones
\cite{KS}, \cite{BGGPG}. Finally, there are recent claims in the literature for
evidence of a neutrino flux histogram \cite{SturrockS} containing two peaks, an
indication of variability pointing towards a nonzero magnetic moment of the neutrino.    

This paper is divided as follows: in section 2 a statistical analysis of the four
rates (Chlorine, Gallium, SuperKamiokande and SNO) on the light of the standard solar model
(BP 2000) \cite{BP2000}  and the most recent data \cite{Homestake,SuperK,SAGE,Gallex+GNO, 
SNO} is presented. The solar neutrino deficit is assumed to be originated from nonstandard 
neutrinos endowed with a magnetic moment interacting with the solar magnetic field. Four 
solar field profiles, all obeying the general features described above, are analysed.
The first three were used in a previous paper \cite{PA} at the time SNO results were
non-existant. Special attention should be payed to profile (4), the one most closely 
following the requirements originated from solar physics investigations \cite{ACT}. All
Gallium measurements (SAGE, Gallex and GNO) are combined in one single data point, so that
the number of d.o.f. is 2: 4 experiments - 2 parameters. These are the mass square 
difference between neutrino flavours $\Delta m^2_{21}$ and $B_{0}$, the value of the
field at the peak, investigated in the ranges $(0-10^{-7})eV^2$ and $(0-300)kG$ 
respectively. This $\Delta m^2$ range is related to neutrino resonances in the region 
from the upper radiation zone to the solar surface, that is, the whole region where a 
significant magnetic field is expected. The upper bound of 300 kG for the peak field 
($B_0$) at the base of the convection zone is suggested by most authors \cite{Parker}, 
\cite{ACT}. The neutrino magnetic moment is taken throughout at $\mu_{\nu}=10^{-11}\mu_B$.
This means that, since the order parameter is the product of $\mu_{\nu}$ by the magnetic 
field, a suitable to fit to solar neutrino data may not allow
a value of $B_0$ well above $10^5G$, as it would imply a value of $\mu_{\nu}$ well
above $3\times10^{-12}\mu_B$, in clear conflict with astrophysical bounds on $\mu_{\nu}$.
\cite{magmo}. The global analysis performed in section 3 (rates + SuperKamiokande spectrum)
therefore excludes all fits with $B_0>10^{5}G$. The border line cases (fits 2.2, 3.1 with 
$B_0=1.18\times10^5G$, $B_0=1.09\times10^5G$) were included. In order to allow for a direct 
comparison with the oscillation cases \cite{KS,BGGPG,BKS1,BGC}, the day/night data from the 
SuperKamiokande collaboration with 1258 days \cite{SuperK} is included in this global 
analysis, despite the fact that no day/night effect is predicted by RSFP. Further,
since the information on the SuperKamiokande total rate is already present in the flux of 
each spectral energy bin, this rate does not enter the analysis, thus following the
same atitude as other authors \cite{KS,BGGPG,BGC}. The number of d.o.f.
is therefore: 3 rates + 19 x 2 spectral points - 2 parameters = 39. The results are 
presented in terms of best fits and allowed physical region (95 \%CL and 99 \%CL)
contours on the $\Delta m^2_{21}$, $B_0$ plane. A small discussion and conclusion section 
then follows.      

\newpage 

\section{Rates only: Chlorine, Gallium, SuperKamiokande and SNO} 

All profiles studied generically satisfy the requirement of a sharp
rise near the bottom of the convective zone with a subsequent smoother 
decrease. The first three were introduced in a previous paper \cite{PA} 
where separate analyses of rates and spectrum were made (no global 
analysis) at a time when SNO results did not exist:

$Profile~(1)~[equilateral~triangle]$
\be
B=0~~~,~~~x<x_R
\ee
\be
B=B_0\frac{x-x_R}{x_C-x_R}~,~x_{R}\leq x\leq x_{C}
\ee
\be
B=B_0\left[1-\frac{x-x_C}{1-x_C}\right]~ ,~x_{C}\leq x \leq 1
\ee
where $x$ is the fraction of the solar radius, $x_R=0.70$, $x_C=0.85$ and all
units are in Gauss. 

$Profile~(2)$
\be
B=0~~~,~~~x<x_R
\ee
\be
B=B_0\frac{x-x_R}{x_C-x_R}~,~x_{R}\leq x\leq x_{C}
\ee
\be
B=B_0\left[1-\left(\frac{x-0.7}{0.3}\right)^2\right]~,~x_{C}<x\leq 1
\ee
with $x_R=0.65$, $x_C=0.75$.

$Profile~(3)$
\be
B=2.16\times10^3~~,~~x\leq 0.7105
\ee
\be
B=B_{1}\left[1-\left(\frac{x-0.75}{0.04}\right)^2\right]~,~0.7105<x<0.7483
\ee
\be
B=\frac{B_{0}}{\cosh30(x-0.7483)}~,~0.7483\leq x\leq 1
\ee
with $B_0=0.998B_1$.

The fourth profile is chosen so as to more closely agree with solar physics observational
requirements \cite{ACT}. In fact a recent detailed study of the solar internal rotation has
lead to the estimation of a large scale magnetic field. An upper limit of 300 000 G
is derived for a field near the base of the convection zone, a value already claimed by
previous authors \cite{Parker} at this particular location. Also a field intensity of
approximately 20 000 G at a depth of 30 000 km below the surface is expected. The
simplest profile satisfying such features is  

$Profile~(4)$
\be
B=0~~,~~x\leq x_R
\ee
\be
B=B_0\frac{x-x_R}{x_C-x_R}~~,~~x_R\leq x \leq x_C
\ee
\be
B=B_0+(x-x_C)\frac{2\times10^{4}-B_0}{0.957-x_C}~~,~~x_C\leq x\leq 0.957
\ee
\be
B=2.10^{4}+(x-0.957)\frac{300-2\times10^{4}}{1-0.957}~~,~~0.957\leq x\leq 1
\ee
with $x_R=0.65$, $x_C=0.713$.

The possibility of this relatively high intensity (20 000 G) at the small depth of 30 000 km 
corresponding to another peak has been known for some time to be inconsistent with the data.
These seem to be consistent only with a field intensity decreasing monotonically 
from the bottom of the convective zone to the surface \cite{PA}.  

The ratios of the RSFP to the SSM event rates $R^{th}_{Ga,Cl}$ are defined as before 
\cite{PA}, the SuperKamiokande one is 
\be
R^{th}_{SK}=\frac{\int_{{E_e}_{min}}^{{E_e}_{max}}dE_e\int_{m_e}^{\infty}dE^{'}\!\!_e
f(E{'}_e,E_e)\int_{E_m}^{E_M}dE\phi(E)[P(E)\frac{d\sigma_{W}}{dT^{'}}+(1-P(E))
\frac{d\sigma_{\bar{W}}}{dT{'}}]}
{\int_{{E_e}_{min}}^{{E_e}_{max}}dE_e\int_{m_e}^{\infty}dE^{'}\!\!_e
f(E{'}_e,E_e)\int_{E_m}^{E_M}dE\phi(E)\frac{d\sigma_{W}}{dT^{'}}}
\ee
%\be
%R_{j}^{th}=\frac{\sum_{i}\int_{{E_e}_j}^{{E_e}_{j+1}}dE_e\int_{{E^{'}_e}_m}^{{E^{'}_e}_M}dE{'}\!\!_e
%f(E{'}_e,E_e)\int_{E_m}^{E_M}dEf_i\phi_{i}(E)[P(E)\frac{d\sigma_{W}}{dT^{'}}+(1-P(E))
%\frac{d\sigma_{\bar{W}}}{dT{'}}]}
%{\sum_{i}\int_{{E_e}_j}^{{E_e}_{j+1}}dE_e\int_{{E^{'}_e}_m}^{{E^{'}_e}_M}dE{'}\!\!_e
%f(E{'}_e,E_e)\int_{E_m}^{E_M}dE\phi_{i}(E)\frac{d\sigma_{W}}{dT^{'}}}
%\ee
Here $\phi(E)$ is the SSM neutrino flux ($hep\!+^8\!\!{B}$) and the
quantity $f(E^{'}_e,E_e)$ is the energy resolution function \cite{Fukuda} of the detector in 
terms of the physical ($E^{'}_e$) and the measured ($E_e$) electron energy ($E_e=T+m_e$). 
The lower limit of $E_{e}$ is the detector threshold energy (${E_e}_{m}={E_e}_{th}$ 
with ${E_e}_{th}$= 5.0MeV) and the upper limit is ${E_e}_{max}=20MeV$ \cite{SuperK}. 
For the lower \cite {PA} and upper \cite{Homepage} integration limits of the neutrino 
energy, one has respectively
\be
E_m=\frac{T^{'}+\sqrt{T{^{'}}^{2}+2m_eT^{'}}}{2}~,~E_{M}=15MeV~(i=^{8}\!\!B)~,~E_{M}=18.8MeV~
(i=hep).
\ee
The weak differential cross sections appearing in equation (17) are given by 
\be
\frac{d\sigma_W}{dT}=\frac{{G_F}^2 m_e}{2\pi}[(g_V+g_A)^2+
(g_V-g_A)^2\left(1-\frac{T}{E}\right)^2-({g_V}^2-{g_A}^2)\frac{m_{e}T}{E^2}]
\ee
for $\nu_{e}e$ scattering, with $g_V=\frac{1}{2}+2sin^2\theta_{W}$, $g_A=\frac{1}{2}$. 
For $\bar\nu_{\mu} e$ and $\bar\nu_{\tau} e$ scattering,
\be
\frac{d\sigma_{\bar{W}}}{dT}=\frac{{G_F}^2 m_e}{2\pi}[(g_V-g_A)^2+
(g_V+g_A)^2\left(1-\frac{T}{E}\right)^2-({g_V}^2-{g_A}^2)\frac{m_e T}{E^2}]
\ee
with $g_V=-\frac{1}{2}+2sin^2\theta_{W}$, $g_A=-\frac{1}{2}$.

The fourth quantity to be evaluated is the ratio of the charged current event rate
by the standard charged current event rate in the SNO experiment ($r_{CC}$). As 
recently announced by the SNO collaboration $r_{CC}=0.347\pm0.029$ \cite{SNO}. By
definition $r_{CC}=R_{CC}/R^{st}_{CC}$ where $R_{CC}$ is given by  
\be
R_{CC}=\int^{\infty}_{T_m}dT_{eff}\int^{\infty}_{Q}dEf(E)P(E)\int^{E+m_{e}-Q}_{m_e}\!\!dE_{e}
\frac{d\sigma_{CC}}{dE_{e}}(E,E_{e})R(E_{{e}_{eff}},E_{e})
\ee
and $R^{st}_{CC}$ stands for the same quantity with the replacement $P(E) \rightarrow 1$. 
In (18) $E_{eff}$ denotes the measured total electron energy $E_{eff}=T_{eff}+m_{e}$ with a
threshold given in terms of the kinetic energy, $T_{m}=6.75MeV$. The function
$R(E_{{e}_{eff}},E_{e})$ is the detector energy response \cite{SNO} and $Q=1.44MeV$ with the
rest of the notation being standard. The CC cross section was taken from K. Kubodera's
homepage, \cite{Kubodera}. 

The partial event rates for each neutrino component in each experiment were 
taken from \cite{BP2000} and the solar neutrino spectra from Bahcall's homepage 
\cite{Homepage}. The contribution of the hep flux to both the Gallium and Homestake 
event rates was neglected. The $\chi^2$ analysis for the ratios of event rates and 
electron spectrum in SuperKamiokande was done following the standard procedure 
described in \cite{PA} \footnote{Here only the main definitions and differences are 
registered. For the calculational details we refer the reader to ref.\cite{PA}.}. The
validity of this procedure and alternative ones for solar neutrinos is discussed in 
refs. \cite {Garz+Giunti}, \cite{CSS}.     

The ratios of event rates to the SSM event rates, both denoted by $R^{th}_j$ in the 
following, were calculated in the parameter ranges $\Delta m^2_{21}=(0-10^{-7})eV^2$, 
$B_0=(0-30)\times10^{4}G$ for all magnetic field profiles and inserted in the $\chi^2$ 
definition, 
\be
\chi^2_{rates}=\sum_{j_{1},j_{2}=1}^{4}({R}^{th}_{j_{1}}-{R_{j_{1}}}^{exp})\left[\sigma^{2}_{rates}
(tot)\right]^{-1}_{j_{1}j_{2}}({R}^{th}_{j_{2}}-{R_{j_{2}}}^{exp})
\ee
with (Ga=1, Cl=2, SK=3, SNO=4). The measured rates ${R_{j}}^{exp}$ in this equation can be 
read directly from table I. Given the four data points and the number of parameters
in the fit, the number of d.o.f. is 2. Within the above parameter ranges for $\Delta m^2_{21}$
and $B_0$, 19 local minima of $\chi^2_{rates}$ were found. They are displayed in table II along
with the respective values of the goodness of fit (g.o.f.). As mentioned in the introduction,
a working value of $\mu_{\nu}=10^{-11}\mu_{B}$ was chosen. Since the order parameter is the
product of the neutrino magnetic moment by the magnetic field which can be as large as
$3\times10^5G$ at the peak, a value $B_0=10^5G$, for example, at the local $\chi^2$ minimum, 
means the existence of a solution with $\mu_{\nu}=\frac{1}{3}10^{-11}\mu_B$. In order to
remain consistent with most astrophysical bounds \cite{magmo}, it is therefore unadvisable
to consider those fits for which $B_0$ is well above $10^{5}G$. A close inspection of table 
II shows that all profiles give reasonable or good quality fits for a peak field value up 
to $\simeq 10^5G$. Based on this criterion and selecting therefore fits 1.1, 1.2, 2.1, 2.2, 
3.1, 4.1 (g.o.f. 34, 34, 38, 25, 42, 51\% respectively) it is seen 
that one obtains comparable fits in RSFP to those in the oscillation cases\cite{KS,BGGPG,BGC}. 
In fact the best oscillation solution from the 'rates-only' analysis is the VAC one \cite{KS} for 
which the g.o.f. is 52\%. \footnote{There is however a minor difference in criterion, since 
these authors considered the SAGE and Gallex/GNO rates as two separate ones, hence 3 d.o.f. 
in their analysis (see table 4 of ref. \cite{KS}).}  Interestingly enough, the best of all fits 
investigated here is 4.1 from profile 4, the one in the closest agreement with solar physics 
observational requirements \cite{ACT}. Whether this remarkable fact is coincidental or essential 
is of course too early to ascertain.

A global analysis of the selected fits 1.1, 1.2, 2.1, 2.2, 3.1 and 4.1 will now be performed 
in section 4.

\section{Global Analysis}

In addition to the rates, the quantity to be investigated now is the SuperKamiokande spectrum
\be
R_{j}^{th}=\frac{\sum_{i}\int_{{E_e}_j}^{{E_e}_{j+1}}dE_e\int_{m_e}^{\infty}dE{'}\!\!_e
f(E{'}_e,E_e)\int_{E_m}^{E_M}dEf_i\phi_{i}(E)[P(E)\frac{d\sigma_{W}}{dT^{'}}+(1-P(E))
\frac{d\sigma_{\bar{W}}}{dT{'}}]}
{\sum_{i}\int_{{E_e}_j}^{{E_e}_{j+1}}dE_e\int_{m_e}^{\infty}dE{'}\!\!_e
f(E{'}_e,E_e)\int_{E_m}^{E_M}dE\phi_{i}(E)\frac{d\sigma_{W}}{dT^{'}}}
\ee
with $j=1,38$.
Following other authors \cite{KS,BGGPG,BGC} and since the information on the SuperKamiokande
total rate is already included in the flux of each spectral energy bin, this rate will now be
disregarded. In order to allow for a straight comparison with other solar neutrino physics
analyses the day/night spectrum is also included although no day/night difference is predicted
by RSFP. Hence the $\chi^2$ definition is as follows
\be
\chi^2_{sp}=\sum_{j_{1},j_{2}=1}^{38}({R}^{th}_{j_1}-R^{exp}_{j_1})[\sigma^{2}_{sp}(tot)]^{-1}
_{j_{1}j_{2}}({R}^{th}_{j_2}-R^{exp}_{j_2})\,.
\ee
The measured rates $R^{exp}_{j}$ $(j=1,...,38)$ in this equation can be directly read from 
table II of ref. \cite{SuperK}.
In performing the global fit one has 41 experiments (3 rates and 38 spectral data) and two
free parameters, namely the mass squared difference between neutrino flavours $\Delta m^2_{21}$
and the peak field value $B_0$, hence 39 d.o.f.. For the global $\chi^2$, 
\be
{\chi^2}_{gl}={\chi^2}_{rates}+{\chi^2}_{spectrum}.
\ee
The local minima of $\chi^2_{gl}$ were investigated for each of the selected four fits from
the previous section (1.1, 1.2, 2.1, 2.2, 3.1, 4.1). The allowed physical regions and stability 
of these fits were next investigated. The analysis proceeds in terms of the two parameters
$\Delta m^2_{21}$, $B_0$ hence two degrees of freedom, and the results are presented in terms 
of 95 and 99\%CL contours ($\Delta \chi^2$=5.99 and 9.21 relative to the local best fit).
Global fits 1.1 and 1.2 of profile 1, whose contours merge, and 3.1 of profile 3 are shown 
together in fig. 1. Global fits 2.1, 2.2 of profile 2, whose contours also merge, are shown in 
fig.2, while 4.1 of profile 4 and its contours is shown in fig. 3. In table III the 
parameter values for each of these global fits including the values of $\chi^2$ and the g.o.f. 
are also shown. It is seen that all g.o.f. are of the same order, with fit 4.1 being, 
interestingly enough, the best, although by a narrow margin. Comparing with the best global fit 
from oscillations (LMA solution) \cite{KS,BGGPG}, the g.o.f. of fit 4.1 is similar but slightly 
smaller (64.7\% compared to 67\% with fixed hep flux) or reasonably smaller (compared 
to 85\% with free hep flux). Since the hep flux was considered fixed in the present 
analysis, only the first comparison should be held as valid. Owing to the small and 
probably not physically meaningful day/night effect (1.29$\sigma$) observed in the SuperKamiokande 
\cite{SuperK}, spectrum and since RSFP predicts no such effect, it is to be expected that 
global fits in RSFP could be marginally worse, although of comparable quality. The
energy spectrum corresponding to global fit 4.1 is shown in fig. 4 superimposed on the
SuperKamiokande data. Allowing for a peak field value $B_0=3\times10^5G$, the
respective value of the magnetic moment, as seen from table III, is 
$\mu_{\nu}=3.2\times10^{-12}\mu_{B}$ with $\Delta m^2_{21}=1.45\times10^{-8}eV^2$. 
 
\section{Concluding Remarks}

The analysis of prospects for the magnetic moment solution to the solar neutrino
problem reveals both rate and global fits of reasonable or good quality. Previous analyses
on RSFP \cite{PA,Valle,GN}, unambiguously indicated from the part of 
the data on rates a preference for solar field profiles
with a steep rise across the bottom of the convection zone in which vicinity they
reach a maximum, followed by a more moderate decrease up to the solar surface.
They showed a quality of rate fits alone which is not shared by the oscillation 
rate fits \cite{KS}. Interestingly enough, this class of profiles is the most consistent one
with solar physics and helioseismology \cite{Parker,ACT}. A further profile (4) was
introduced in the present paper which is the simplest most closely following the
present requirements from solar physics \cite{ACT}. It is important to note that,
as found in the present paper, this is the profile giving the best of all RSFP fits, 
both for the rates and global. For this (global) fit one finds that 
$\Delta m^2_{21}=1.45\times10^{-8}eV^2$ and $\mu_{\nu}=3.2\times10^{-12}\mu_{B}$, 
(g.o.f. 64.7\%) if one allows for a solar field at the bottom of the convective zone 
$B_0=300~000G$. An analysis of spin flavour precession, both resonant and non-resonant, 
was also performed in a recent paper \cite{Miranda}, whose authors limit themselves to 
a 'rates-only' analysis with a single profile. They find a better g.o.f. for the 
resonant (39\%) than for the non-resonant case (28\%), however substantially smaller 
than the best one obtained here with profile 4.   
  
The forthcoming neutrino experiments Borexino \cite{Borexino} and Kamland \cite{Kamland} 
will be crucial to test the validity of the RSFP hypothesis against oscillations. In fact,
at the present stage both the LMA and RSFP solutions give global fits of equally good 
quality and, as far as rates only are concerned, RSFP solutions provide the best of
all fits. Furthermore, owing to the typical shape of the RSFP survival probability \cite{PA}, 
which is quite different from the LMA, LOW and VAC ones at the intermediate energies 
\cite{BKS}, the RSFP solution predicts for Borexino a clearly smaller neutrino-electron 
scattering \cite{AP} event rate than these three oscillation solutions whose prediction 
is essentially the same \cite{BGGPG}. It is quite likely that the
Kamland experiment will either bring confirmation of the LMA scenario, or either disprove
it, along with the less favoured solutions, LOW and VAC \cite{BGGPG}. Consequently, in case
these are excluded by Kamland, the expectation for Borexino will naturally be that of a
strong reduction relative to the standard event rate, therefore providing evidence for 
RSFP, unless a clear day/night effect is observed.  

In the present analysis only time averaged data were considered and a fitting was made 
to a time constant profile 'buried' in the solar interior. If, on the contrary, the active
neutrino flux turns out to be time dependent, a situation most likely to be interpreted 
through the magnetic moment solution with a time dependent interior field, the present 
approach is obviously inadequate. Averaging the event rates over time implies disregarding 
possible information in the data which otherwise is available if different periods of time 
are considered \cite{SturrockS}. The robustness of such a procedure will greatly improve 
with the accumulation of more data.

\newpage

%\centerline{\large Figure captions}

\begin{center}
\begin{tabular}{lcccc} \\ \hline \hline
Experiment &  Data      &   Theory   &   Data/Theory  &  Reference \\ \hline
Homestake  &  $2.56\pm0.16\pm0.15$ & $7.7\pm^{1.3}_{1.1}$ & $0.332\pm0.05$ & 
\cite{Homestake} \\
Ga     &  $74.7\pm5.13$ & $129\pm ^8_6$ & $0.58\pm0.06$ &
\cite{Gallex+GNO},\cite{SAGE} \\
SuperKamiokande&$2.32\pm0.085$ &
$5.05\pm^{1.0}_{0.7}$&$0.459\pm0.005\pm^{0.016}_{0.018}$& \cite{SuperK}\\ 
SNO&$1.75\pm{0.15}$&$5.05\pm^{1.0}_{0.7}$&$0.347\pm0.029$& \cite{SNO} \\ \hline
\end{tabular}
\end{center}

{Table I - Data from the solar neutrino experiments. Units are SNU for
Homestake and Gallium and $10^{6}cm^{-2}s^{-1}$ for SuperKamiokande and SNO. The
result for Gallium is the combined one from SAGE and Gallex+GNO.}

\begin{center}
\begin{tabular}{ccccc}\\ \hline \hline
Fit & $\Delta m^2_{21}(eV^2)$ & $B_0~(G)$ & $\chi^2_{rates}$/2 d.o.f. & g.o.f.\\ \hline
1.1        & $7.04\times10^{-9}$ & $4.2\times10^{4}$ & $2.15$ & $34.1\%$\\
1.2        & $7.35\times10^{-9}$ & $6.5\times10^{4}$  &$2.17$ & $33.8\%$\\
1.3        & $8.38\times10^{-9}$ & $1.35\times10^{5}$ &$1.32$ & $51.7\%$\\ 
1.4        & $8.62\times10^{-9}$ & $1.67\times10^{5}$ & $1.81$ & $40.4\%$ \\
1.5        & $1.0\times10^{-8}$ & $2.34\times10^{5}$ & $1.12$ &$57.1\%$  \\ \hline
2.1        & $1.28\times10^{-8}$ & $9.6\times10^{4}$ & $1.96$ & $37.6\%$\\
2.2        & $1.29\times10^{-8}$ & $1.18\times10^{5}$ & $2.75$ & $25.3\%$\\
2.3        & $1.37\times10^{-8}$ & $1.72\times10^{5}$  &$1.51$  & $47.1\%$\\
2.4        & $1.38\times10^{-8}$ & $1.96\times10^{5}$  &$2.37$  & $30.6\%$\\ 
2.5        & $1.43\times10^{-8}$ & $2.47\times10^{5}$  & $1.12$  &$57.3\%$ \\
2.6        & $1.48\times10^{-8}$ & $2.73\times10^{5}$  & $2.19$  &$33.5\%$ \\ 
2.7        & $1.59\times10^{-8}$ & $3.23\times10^{5}$  & $0.986$  &$61.1\%$ \\ \hline
3.1        & $1.45\times10^{-8}$ & $1.09\times10^{5}$ & $1.73$ & $42.2\%$\\
3.2        & $1.87\times10^{-8}$ & $1.83\times10^{5}$  &$2.15$  & $34.1\%$\\ \hline
4.1        & $1.43\times10^{-8}$ & $9.8\times10^{4}$  &$1.34$  & $51.2\%$\\ 
4.2        & $1.41\times10^{-8}$ & $1.31\times10^{5}$  &$3.41$  & $18.2\%$\\ 
4.3        & $1.49\times10^{-8}$ & $2.04\times10^{5}$  &$1.17$  & $55.8\%$\\ 
4.4        & $1.49\times10^{-8}$ & $2.38\times10^{5}$  &$2.74$  & $25.4\%$\\ 
4.5        & $1.58\times10^{-8}$ & $3.05\times10^{5}$  &$1.10$  & $57.8\%$\\ \hline
\end{tabular}
\end{center}  

{Table II - Rate fits: local minima of $\chi^2_{rates}$ for profiles 1, 2, 3, 4
(eqs.(1)-(13)). Only the fits 1.1, 1.2 (profile 1), 2.1, 2.2 (profile 2), 3.1 
(profile 3) and 4.1 (profile 4) are considered for the global analysis (rates + 
day/night spectrum). The remainder correspond to a value of $B_0$ largely 
exceeding $10^5G$. See the text for details.}

\newpage

\begin{center}
\begin{tabular}{ccccc}\\ \hline \hline
Fit & $\Delta m^2_{21} (eV^2)$ & $B_0(G)$ & $\chi^2_{gl}/39 d.o.f.$ & g.o.f.\\ \hline
1.1   & $6.80\times10^{-9}$ & $4.0\times10^{4}$ & 36.2 & 59.7\%  \\
1.2  & $7.07\times10^{-9} $ & $6.8\times10^{4}$ & 36.3 & 59.2\%   \\
2.1  & $1.27\times10^{-8} $ & $9.4\times10^{4}$ & 36.0 & 60.9\%   \\
2.2  & $1.26\times10^{-8} $ & $1.22\times10^{5}$ & 38.4 & 49.5\%   \\
3.1 & $1.41\times10^{-8} $ & $1.04\times10^{5}$ & 35.4 & 63.6\%   \\
4.1  & $1.45\times10^{-8} $ & $9.6\times10^{4}$ & 35.1 & 64.7\%  \\  \hline
\end{tabular}
\end{center}

{Table III - Global fits with the three rates (Homestake, Gallium, SNO) and the 
SuperKamiokande day/night spectrum. Fit 4.1, besides being the best, although by 
a small margin, is related to a solar field profile which agrees most closely with
the requirements from a solar physics analysis (details in the text).}

\vspace{1cm} 

%\newpage

\centerline{\large Figure captions}

%\vglue 0.4cm
%\vspace{0.2cm} 
\noindent
Fig. 1.
Global fits 1.1, 1.2 (lower left, corresponding to profile (1)) and 3.1 (upper
right, corresponding to profile 3) and their 95\% (dashed) and 99\% CL (solid) 
contours. See also table III. 

\noindent
Fig. 2.
Same as fig.1 for global fit 2.1, 2.2 (profile 2). See also table III. 

\noindent
Fig. 3.
Same as fig.1 for global fit 4.1 (profile 4). See also table III. 

\noindent
Fig. 4.
The theoretical prediction of the SuperKamiokande recoil electron spectrum (solid line)
superimposed on the data set \cite{SuperK} (1258 days) for global fit 4.1 (profile 4). 
This is also the best fit found in the analysis, for which $\chi^2_{global}=35.1$  
(or g.o.f.=64.7\%) for 39 d.o.f..

\newpage

\begin{figure}
%\begin{picture}(18,17)
%%%\put(0,1){\epsfig{figure=fig1.eps,width=17cm}}
\mbox{\psfig{figure=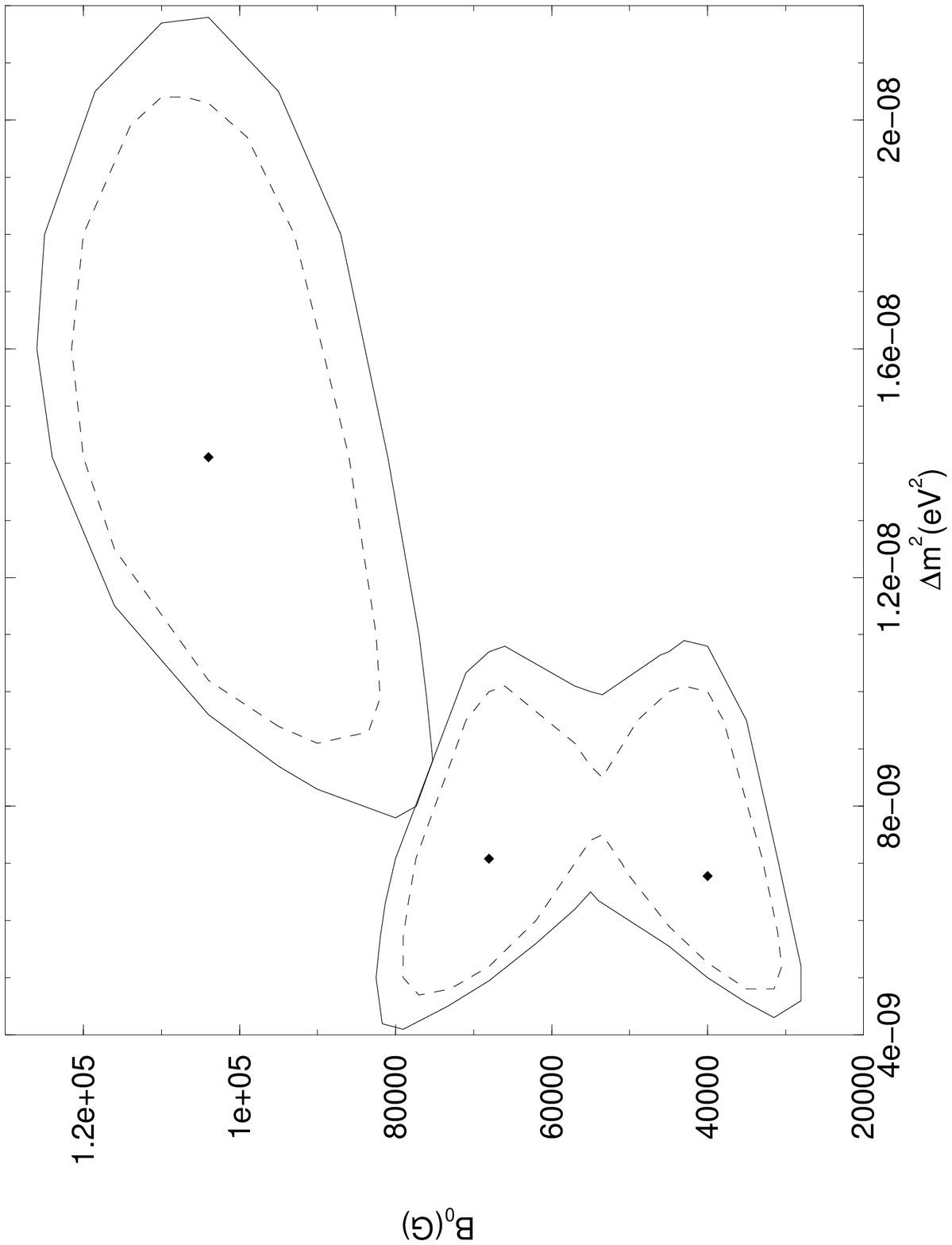,width=15cm}}
%\end{picture}
%\caption{Equilateral triangle field profile given by eqs. (1) - (3) as a function
%of the solar coordinate $x=r/R_{S}$ with $x_{R}=0.7$, $x_{C}=0.85$ . Field strength
%units are in Gauss.}
\centerline{\mbox{Fig. 1.}}
\end{figure}

\begin{figure}
%\begin{picture}(18,17)
%%%\put(0,1){\epsfig{figure=fig1.eps,width=17cm}}
\mbox{\psfig{figure=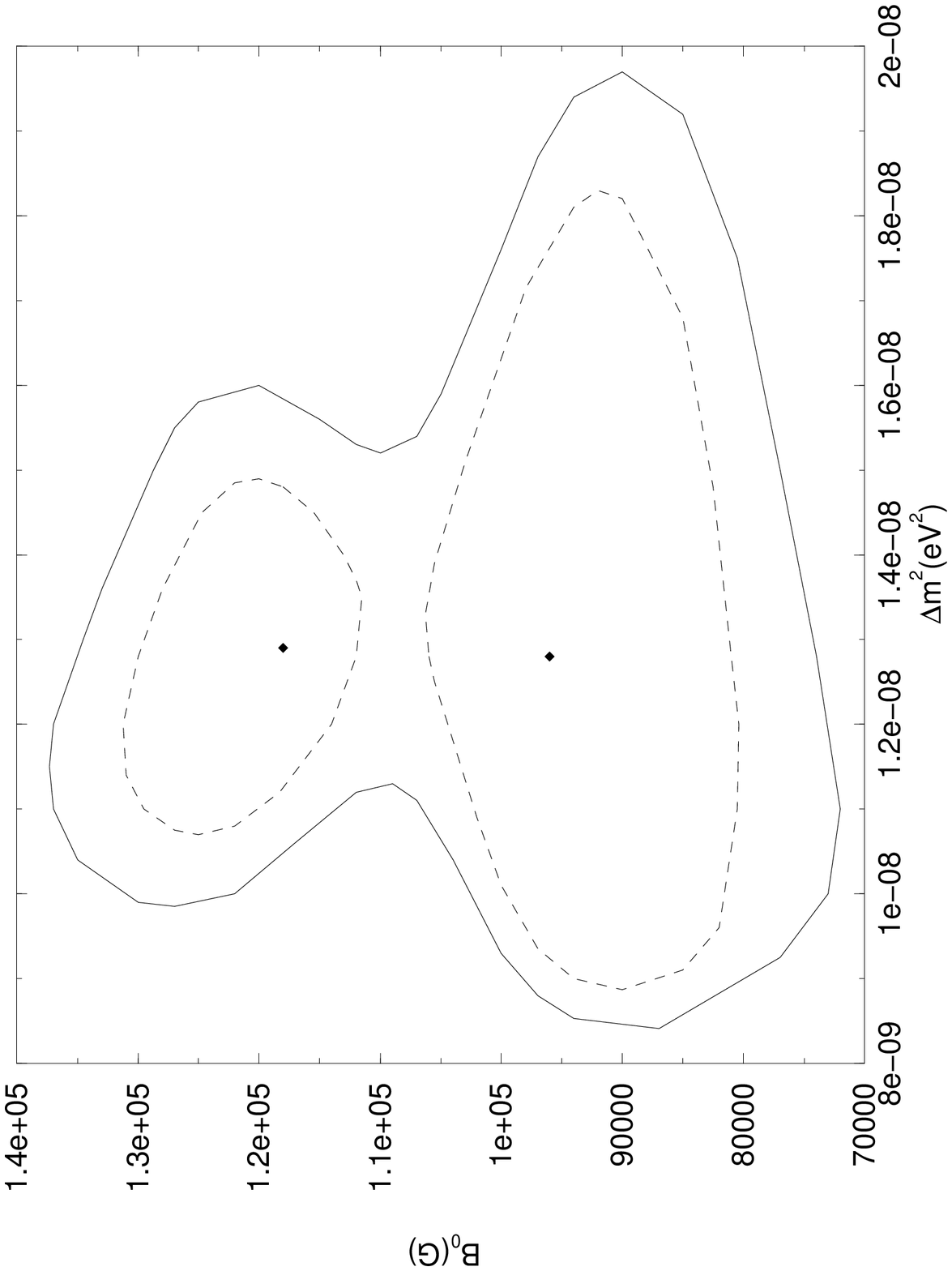,width=15cm}}
%\end{picture}
%\caption{Equilateral triangle field profile given by eqs. (1) - (3) as a function
%of the solar coordinate $x=r/R_{S}$ with $x_{R}=0.7$, $x_{C}=0.85$ . Field strength
%units are in Gauss.}
\centerline{\mbox{Fig. 2.}}
\end{figure}

\newpage

\begin{figure}
%\begin{picture}(18,17)
%%%\put(0,1){\epsfig{figure=fig1.eps,width=17cm}}
\mbox{\psfig{figure=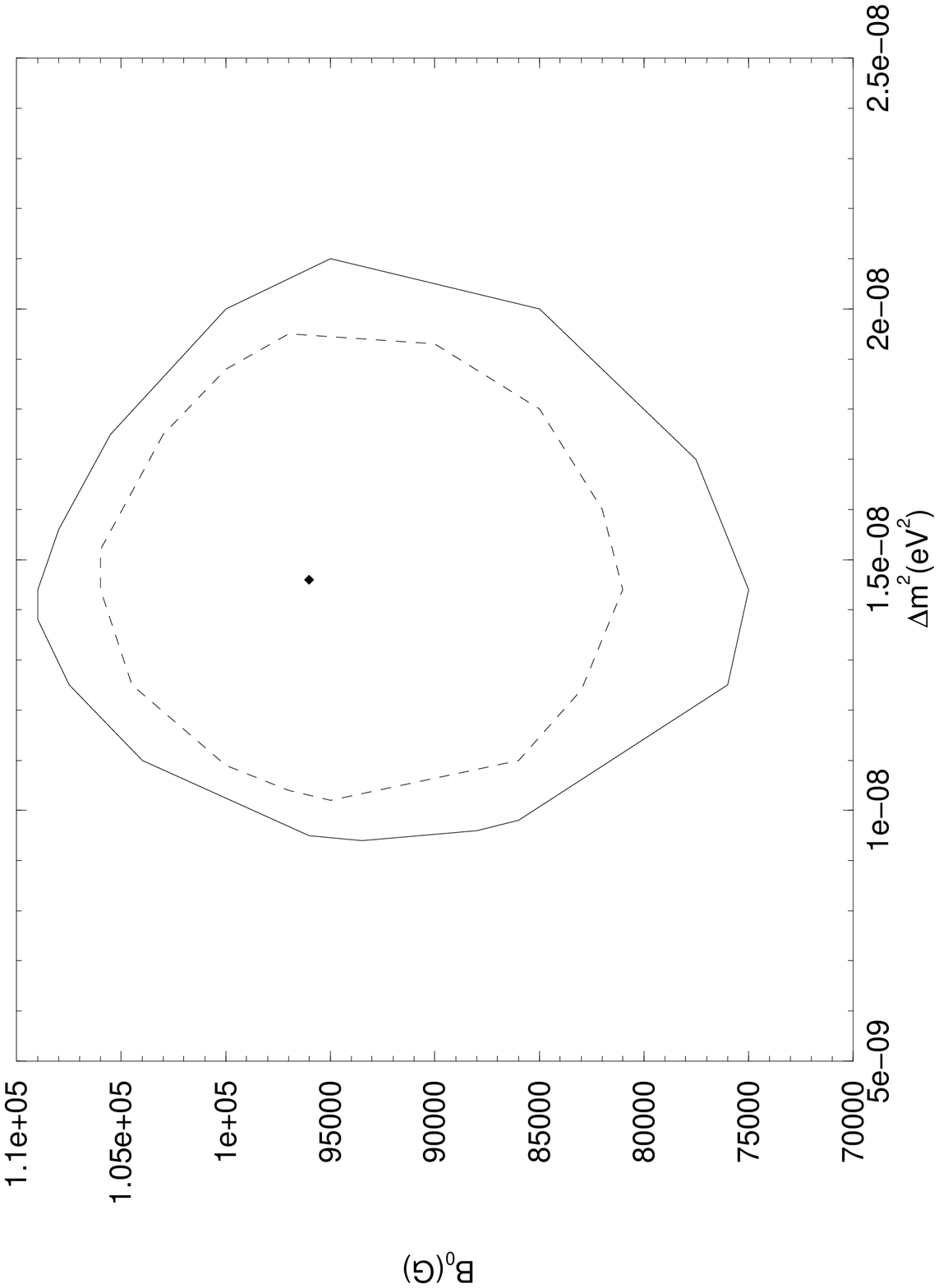,width=15cm}}
%\end{picture}
%\caption{Equilateral triangle field profile given by eqs. (1) - (3) as a function
%of the solar coordinate $x=r/R_{S}$ with $x_{R}=0.7$, $x_{C}=0.85$ . Field strength
%units are in Gauss.}
\centerline{\mbox{Fig. 3.}}
\end{figure}

\newpage

\begin{figure}
%\begin{picture}(18,17)
%%%\put(0,1){\epsfig{figure=fig1.eps,width=17cm}}
\mbox{\psfig{figure=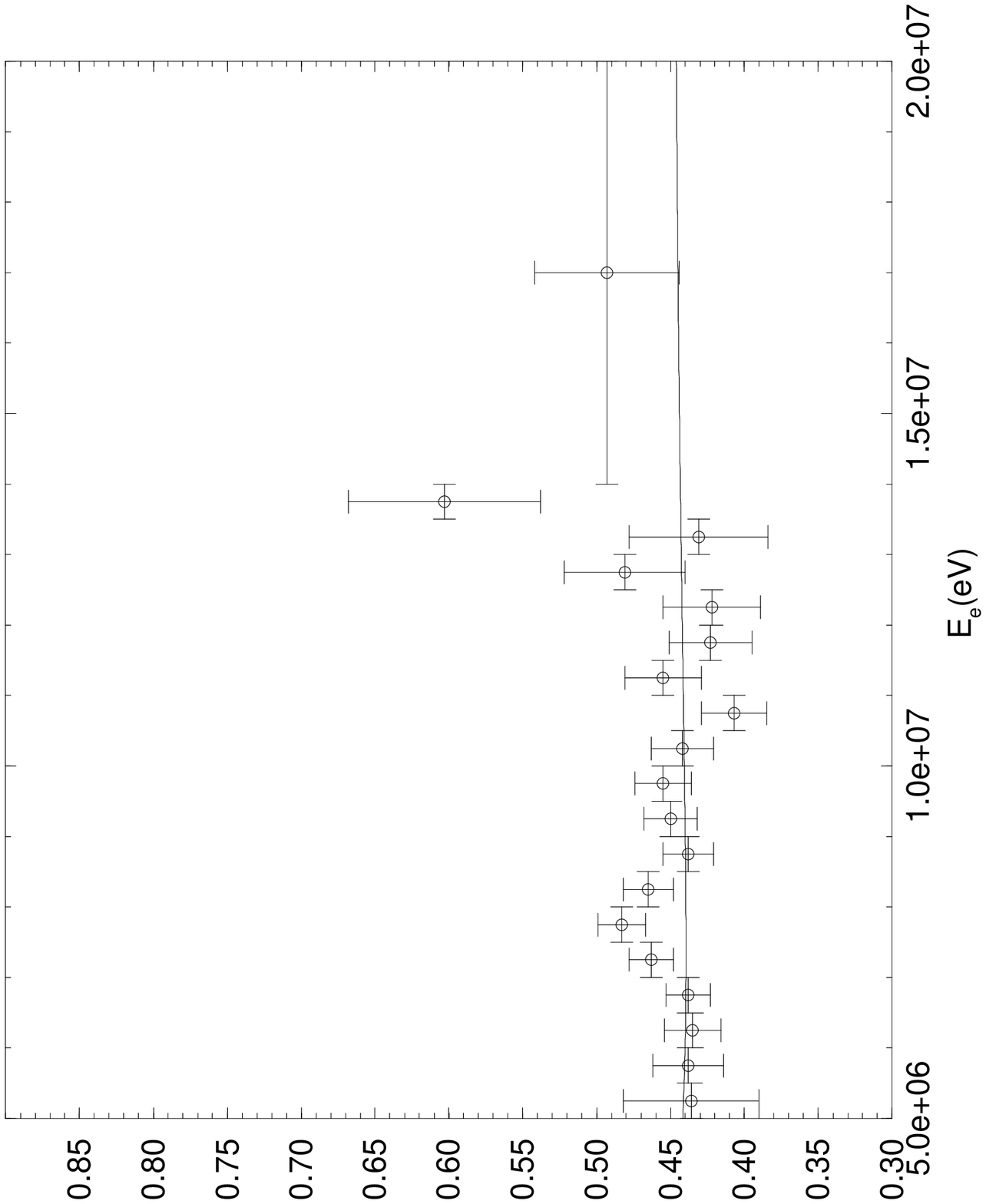,width=15cm}}
%\end{picture}
%\caption{Equilateral triangle field profile given by eqs. (1) - (3) as a function
%of the solar coordinate $x=r/R_{S}$ with $x_{R}=0.7$, $x_{C}=0.85$ . Field strength
%units are in Gauss.}
\centerline{\mbox{Fig. 4.}}
\end{figure}

\end{document}